\documentstyle[12pt]{article}
\begin{document}
\baselineskip=14pt

\begin{center}
{\Large {\bf A Dirac-type equation for spacelike
 neutrinos}}

\vskip 0.8cm
            Tsao Chang \\ 
  Center for Space Plasma and Aeronomy Research\\
  University of Alabama in Huntsville\\
          Huntsville, AL 35899\\
      Email: changt@cspar.uah.edu\\

\vskip 0.5cm
\end{center}

\baselineskip=14pt

  Based on experimental evidences supporting the hypothesis that 
neutrinos might be spacelike particles, a new Dirac-type equation 
is proposed and a spin-$\frac{1}{2}$ spacelike quantum theory is 
developed.  The new Dirac-type equation provides a solution for 
the puzzle of negative mass-square of neutrinos.  This equation 
can be written in two spinor equations coupled together via 
nonzero mass while respecting maximum parity violation, and 
it reduces to one Weyl equation in the massless limit. Some 
peculiar features of spacelike neutrino are discussed in this 
theoretical framework.


\newpage
\noindent
{\bf I. INTRODUCTION}

 A model has recently been presented to fit the cosmic ray 
spectrum at $E \approx 1-4$ PeV [1] using the hypothesis
that the electron neutrino is a tachyon.  This model yields 
a value for $m^2(\nu_e) \approx -3 \ eV^2$, which 
is consistent with the results from recent 
measurements in tritium beta decay experiments [2-4]. Moreover,
the muon neutrino also exhibits a negative mass-square [5]. 
However, up to now, there is no satisfactory relativistic quantum
theory to describe neutrinos as spin-$\frac{1}{2}$ tachyons.  

   The negative value of the neutrino mass-square simply means:
     $$  E^2/c^2 - p^2 = m_{\nu}^2 c^2 < 0	\eqno  (1)  $$    
The right-hand side in Eq.(1) can be rewritten as ($-m_s^2 c^2$), 
then $m_s$ has a positive value. The subscript $s$ means
spacelike particle, i.e. tachyon. 

 Based on special relativity and known as re-interpretation rule, 
tachyon as a hypothetical particle was proposed by Bilaniuk et al. 
in the sixties [6-8].  For tachyons, the relation of momentum 
and energy is shown in Eq.(1). The negative value on the right-
hand side of Eq.(1) means that $p^2$ is greater than $(E/c)^2$.
The velocity of a tachyons, $u_s$, is greater than speed of light. 
The momentum and energy in terms of $u_s$ are as follows:
    $$	p =  \frac{m_s u_s}{\sqrt{u_s ^2 /c ^2  - 1}},  \quad 		
    	E = \frac{m_s c^2}{\sqrt{u_s ^2 /c ^2  - 1}}     \eqno  (2) 
$$	

  Any physical reference system is built by timelike particles 
(such as atoms, molecules etc.), which requires that a reference 
frame must move slower than light.  On the other hand, once a tachyon 
is created in an interaction, its speed is always greater than 
the speed of light.  The neutrino is the most likely candidate for a 
tachyon because it has left-handed spin in any reference 
frame [9,10].  However, anti-neutrino always has right-handed 
spin.  Considering the measured mass-square is negative
for the muon neutrino,  Chodos, Hauser and Kostelecky [9] suggested 
in 1985 that the muon neutrino might be a tachyon.  They also 
suggested that one could test the tachyonic neutrino in high energy 
region using a strange feature of tachyon: $E_\nu$ could be negative 
in some reference frames [11,12].  This feature has been further 
studied by Ehrlich [1].  Therefore, it is required to construct a 
spacelike quantum theory for neutrinos.       

The first step in this direction is usually to introduce an 
imaginary mass, but these efforts could not reach the point of 
constructing a consistent quantum theory.  Some early investigations
of Dirac-type equations for tachyons are listed in 
Ref. [13,14].  An alternative approach was investigated by Chodos et al.
[9].  They examined the possibility that muon neutrino might be
tachyonic fermion.  A form of the lagrangian density for tachyonic 
neutrinos was proposed.  Although they did not obtain a satisfactory 
quantum theory for tachyonic fermions, they suggested that more 
theoretical work would be needed to determine a physically acceptable 
theory.  
\\

\noindent
{\bf II. A NEW DIRAC-TYPE EQUATION}

   In this paper, we will start with a different approach for deriving 
a new Dirac-type equation for spacelike neutrinos.  In order to 
avoid introducing imaginary mass, Eq. (1) can be rewritten as
    $$ E = (c^2p^2 -  m_s^2c^4 )^{1/2}  \eqno  (3)  $$
where $m_s$ is called proper mass.  For instance, 
$m_s(\nu_e)$= 1.6 eV, if taking $m^2(\nu_e)= -2.5 \ eV^2$ [15].  
To follow Dirac's approach [16], the Hamiltonian must be first 
order in the momentum operator ${\hat p}$:
  $$  \hat E = c {\vec \alpha} \cdot {\hat p} + \beta_s m_s c^2 	
     \eqno (4)  $$
with  ($\hat E = i\hbar \partial /\partial t , {\hat p} = 
-i \hbar \nabla $).  ${\vec \alpha} = (\alpha_1, \alpha_2, 
\alpha_3$) and $\beta_s$ are 4$\times$ 4 matrix, which are defined as
  $$ {\alpha_i} = \left(\matrix{0&\sigma_i\cr
                         \sigma_i&0\cr}\right),  \quad
   \beta_s = \left(\matrix{0&I\cr
                         -I&0\cr}\right)  \eqno (5)   $$
where $\sigma_i$ is 2$\times$2 Pauli matrix, $I$ is 2$\times$2 unit 
matrix.  Notice that $\beta_s$ is a new matrix, which 
is different from the one in the traditional Dirac equation.  
We will discuss the property of $\beta_s$  in a later section.

    When we take the square of both sides of Eq. (4), and consider the 
following relations:
 $$ \alpha_i \alpha_j + \alpha_j \alpha_i = 2 \delta_{ij}    $$
      $$	\alpha_i \beta_s + \beta_s \alpha_i =  0  $$
      $$       \beta_s^2 = -1	\eqno    (6)		$$
then the relation in Eq.(1) or Eq. (3) is reproduced.  Since Eq.(3) is 
related to Eq. (2), this means $\beta_s$ is the right choice to 
describe neutrinos as tachyons.  Notice that the relation between 
the matrix $\beta_s$ and the traditional matrix $\beta$ is as follows:
$$  \beta_s = \beta \gamma_5 \,,   \quad where \quad
   {\beta} = \left(\matrix{I&0 \cr   0&-I \cr}\right) \,, \quad  
  {\gamma_5} = \left(\matrix{0&I \cr I&0 \cr}\right)
  \eqno	(7)   $$ 

 We now study the spin-$\frac{1}{2}$ the property of neutrino (or
anti-neutrino) as a tachyonic fermion.  

  Denote the wave function as
  $$ \Psi = \left(\matrix{\varphi ({\vec x},t)\cr
                         \chi ({\vec x},t)\cr}\right) \quad
with \quad
   \varphi = \left(\matrix{\varphi_1\cr
                         \varphi_2\cr}\right), \quad
    \chi = \left(\matrix{\chi_1\cr
                         \chi_2\cr}\right)    $$
the complete form of the new Dirac-type equation, Eq. (4), becomes
$$  \hat E \Psi = c({\vec \alpha} \cdot {\hat p})\Psi + 
           \beta_s m_s c^2 \Psi 	     \eqno (8)  $$
It can also be rewritten as a pair of two-component equations:
$$ i\hbar \frac{\partial \varphi}{\partial t} = -ic 
    \hbar {\vec \sigma} \cdot  \nabla \chi + m_s c^2 \chi   $$
	$$ i\hbar \frac{\partial \chi}{\partial t} = -ic \hbar 
\vec{\sigma} \cdot \nabla \varphi - m_s c^2 \varphi  
\eqno (9)   $$

From the equation (8), the continuity equation is derived:
$$  \frac{\partial \rho}{\partial t} +
           \nabla \cdot {\vec j} = 0  \eqno (10)   $$
and we have
Eq.(10) can be rewritten as
  $$ \rho = \Psi^{\dag}  \gamma_5 \Psi ,  \quad
  {\vec j} = c(\Psi^{\dag} \gamma_5 {\vec \alpha} \Psi) 
     \eqno (11)  $$
where $\rho$ and $\vec j$ are probability density and current; 
$\Psi^{\dag}$ is the Hermitian adjoint of $\Psi$ .

Considering a plane wave along the $z$ axis for a right-handed 
particle, the helicity $H = ({\vec \sigma} \cdot {\vec p})/|{\vec p}|
 = 1 $, then Eq. (8) or (9) yields the following relation:
  $$	 \chi = \frac{cp - m_s c^2}{E} \varphi   \eqno	(12) $$
\\

\noindent
{\bf III. COVARIANT FORM AND EXPLICIT SOLUTIONS}

The new Dirac-type equation (8) can be written in a covariant form:
$$   i {\hbar} \gamma^{\mu}\partial_{\mu} \Psi- m_s c
   \gamma_5 \Psi  = 0    \eqno (13)  $$  
Here the standard convention for the Dirac matrices are used:
$$ \gamma^0 ={\beta} = \left(\matrix{I&0 \cr   0&-I \cr}\right) \quad  
  {\gamma^i} = \left(\matrix{0&\sigma_i \cr -\sigma_i&0 \cr}\right)
  \eqno   (14)   $$ 
For a free particle with momentum $\vec p$ in the $z$ direction,
the plane wave can be represented by
 $$ \Psi(z,t)=\psi_{\sigma}exp[\frac{i}{\hbar}(pz-Et)]  
\eqno (15)   $$
where $\psi_{\sigma}$ is a four-component bispinor.  Substituting
this bispinor into the wave equation (8) or (13), the explicit form
of two bispinors with positive-energy states are listed as follows:
$$ \psi_1=\psi_{\uparrow (+)} = N \left(\matrix{1\cr
     0\cr A \cr 0 \cr}\right), \quad
    \psi_2= \psi_{\downarrow (+)}  = N \left(\matrix{0\cr
                       -A \cr 0 \cr 1 \cr}\right)    \eqno (16) $$
and other two bispinors with the negative-energy states are:
$$ \psi_3=\psi_{\uparrow (-)} = N \left(\matrix{1\cr
     0 \cr -A \cr 0 \cr}\right), \quad
    \psi_4= \psi_{\downarrow (-)}  = N \left(\matrix{0\cr
                       A \cr 0 \cr 1 \cr}\right)    \eqno (17) $$
where the component $A$ and the normalization factor $N$ are
$$  A=\frac{cp-m_s c^2}{|E|},  \quad
    N=\sqrt{\frac{p+m_s c}{2m_s c}}        \eqno (18)  $$
For $ \psi_1=\psi_{\uparrow (+)}$, the conserved current in Eq.(11)
becomes:
$$ \rho = \Psi_1{^\dag} \gamma_5 \Psi_1=\frac{|E|}{m_s c^2}, \quad
     j = \frac{p}{m_s}     \eqno (19)  $$

Let ${\bar\Psi} = \Psi^{\dag} \beta $, we can obtain the following scalars:
$$  {\bar\Psi_1} \Psi_1  = {\bar\Psi_3} \Psi_3  = 1 
   \eqno (20a)  $$
$$  {\bar\Psi_2} \Psi_2  = {\bar\Psi_4} \Psi_4  = -1 
   \eqno (20b)  $$
In addition, the pseudo scalar for each spinor satisfies:
$$  {\bar\Psi} \gamma_5 \Psi  = 0   \eqno (21)  $$

\noindent
{\bf IV.  PARITY VIOLATION FOR NEUTRINOS}
  
 In order to compare the new Dirac-type equation Eq.(7) with
the two component Weyl equation in the massless limit, 
we now consider a linear combination of $\varphi$ and $\chi$ :
  $$   \xi= {1 \over {\sqrt 2}} (\varphi + \chi)  , \quad
      \eta = {1 \over {\sqrt 2}} (\varphi - \chi)  \eqno (22)  $$
where $\xi ({\vec x},t)$ and $\eta ({\vec x},t)$ are two-component 
spinor functions. In terms of $\xi$ and $\eta$, Eq.(10) becomes 
  $$ \rho = \xi^{\dag} \xi - \eta^{\dag} \eta ,  \quad
 {\vec j} = c(\xi^{\dag} {\vec\sigma} \xi + \eta^{\dag}{\vec \sigma}
 \eta) \eqno (23)  $$
Moreover, Eq.(8) can be rewritten in Weyl representation: 
$$ i\hbar \frac{\partial \xi}{\partial t} = -ic \hbar {\vec \sigma}
    \cdot \nabla \xi - m_s c^2 \eta   $$
$$ i\hbar \frac{\partial \eta}{\partial t} = ic \hbar {\vec \sigma}
     \cdot  \nabla \eta + m_s c^2 \xi  \eqno (24)   $$
In the above equations, both $\xi$ and $\eta$ are coupled via 
the mass term $m_s$.

  For comparing Eq. (24) with the well known Weyl equation, 
we take a limit, $m_s = 0$, then the first equation in
Eq. (24) reduces to
	$$   \frac{\partial \xi_{\bar{\nu}}}{\partial t} = -c
 {\vec \sigma} \cdot \nabla \xi_{\bar{\nu}}     \eqno (25)   $$
In addition, the second equation in Eq. (24) vanishes because 
$\varphi = \chi$ when $m_s = 0$.

Eq. (25) is the two-component Weyl equation for describing 
anti-neutrinos ${\bar{\nu}}$, which is related to the 
maximum parity violation discovered in 1956 
by Lee and Yang [18,19].  They pointed out that no experiment had 
shown parity to be a good symmetry for weak interactions.  Now we see 
that, in terms of Eq.(24), once if neutrino has some mass, no 
matter how small it is, two equations should be coupled together via 
the mass term while still respecting maximum parity violation. 

 Indeed, the Weyl equation (25) is only valid for antineutrinos 
since a neutrino always has left-handed spin, which is opposite 
to antineutrino.  For this purpose, we now introduce a 
transformation:
        $$   \vec{\alpha}  \rightarrow - {\vec \alpha}
 	\eqno (26) $$
It is easily seen that the anticommutation relations in Eq. (6) remain 
unchanged under this transformation.  In terms of Eq. (26), Eq.(4) 
becomes
$$  \hat E \Psi_\nu = -c ({\vec \alpha} \cdot {\hat p})\Psi_\nu + 
           \beta_s m_s c^2 \Psi_\nu 	     \eqno (27)  $$
Furthermore, ${\vec \sigma}$ should be replaced by
(${-\vec \sigma}$) from Eq. (5-9) and Eq.(24) for describing a neutrino.
Therefore, the two-component Weyl equation for massless neutrino 
becomes: 
 $$   \frac{\partial \xi_\nu}{\partial t} = c{\vec \sigma} \cdot
           \nabla \xi_\nu      \eqno (28)   $$
 In fact, the transformation (26) is associated with the CPT theorem.
Some related discussions can be found in Ref.[20,21].\\  

\noindent
{\bf V. REMARKS}

 In this paper, The hypothesis that neutrinos might be tachyons
is further investigated. A spin-${\frac{1}{2}}$ spacelike quantum 
theory is developed on the basis of the new Dirac-type equation. 
It provides a solution for the puzzle of negative mass-square 
of neutrinos.
 
   Spacelike neutrinos have many peculiar features, which 
are very different from all other particles.  For instance,  
neutrinos only have weak interactions with other particles.
Neutrino has left-handed spin in any reference 
frame.  On the other hand, anti-neutrino always has 
right-handed spin.  This means that the speed of neutrinos
must be equal to or greater than the speed of light. Otherwise,
the spin direction of neutrino would be changed in some
reference frames.  Moreover, the energy of a tachyonic
neutrino (or anti-neutrino), $E_\nu$, could be negative in 
some reference frames.  We will discuss the subject of the 
negative energy in another paper. 

 The electron neutrino and the muon neutrino may have different
non-zero proper masses.  If taking the data from Ref.[15], then 
we obtain $m_s(\nu_e)=1.6 \ eV$ and $m_s(\nu_{\mu})=0.13 \ MeV$.
In this way, we can get a natural explanation why the numbers 
of e-lepton and $\mu$-lepton are conserved respectively.

 Comparing with the electron mass, the mass term of the e-neutrino
in Eq.(13) is approximately close to zero.  Moreover, from Eq.(21),
${\bar\Psi} \gamma_5 \Psi  = 0$ for Spacelike neutrinos.  It means
that the mass term in Eq.(13) may be negligible in most cases.  
In fact, the momentum of a neutrino is much greater than $m_s c$ in 
most measurements. For instance, let $p_s = 10 m_s c = 16 eV/c$,
Eq.(2) yields the speed of e-neutrino: $u_s = 1.005 c$. Therefore, 
spacelike neutrinos behave just like the massless neutrinos.  
Besides, we have the coefficient $A \simeq 1$ in Eq.(18). This 
 similarity may also play role at the level of quantum field theory 
and SU(2) gauge theory.

 According to special relativity [22], if there is a spacelike 
particle, it might travel backward in time.  However, a re-
interpretation rule has been introduced since the sixties [6-8].  
Another approach is to introduce a kinematic time under a non-
standard form of the Lorentz transformation [23-27].  Therefore, 
special relativity can be extended to the spacelike region, and 
tachyons are allowed without causality violation.

   Generally speaking, the above spin-${\frac{1}{2}}$ spacelike 
quantum theory provides a theoretical framework to study the
hypothesis that neutrinos are tachyonic fermions.  More
measurements on the cosmic ray at the spectrum knee and more 
accurate tritium beta decay experiments are needed to further 
test the above theory.\\

\vskip 0.6cm
  The author is grateful to G-j. Ni and Y. Takahashi 
for helpful discussions.  \\


\baselineskip=14pt
\noindent
\begin{thebibliography}{99}

\bibitem{1} R. Ehrlich, Phys. Lett. B {\bf 493}, (2000) 229;
 Phys. Rev. D {\bf 60}, (1999) 17302;  Phys. Rev. D {\bf 60}, 
(1999) 73005.

\bibitem{2} Ch. Weinhermer et al., Phys. Lett. B {\bf 460} 
(1999) 219.

\bibitem{3} V. M. Lobashev et al., Phys. Lett. B {\bf 460} 
(1999) 227.

\bibitem{4} J. Bonn and Ch. Weinheimer, Acta Phys. Pol., {\bf 31}
(2000) 1209.

\bibitem{5} K. Assamagan et al., Phys. Rev. D {\bf 53} 
(1996) 6065.

\bibitem{6} O.M.P. Bilaniuk, V.K. Deshpande, and E.C.G. 
Sudarshan, Am.J.Phys. {\bf 30} (1962) 718.

\bibitem{7} E. Recami et al, Tachyons, Monopoles and Related 
  Topics, North-Holland, (1978), and references therein.

\bibitem{8} G. Feinberg, Phys. Rev. {\bf 159} (1967) 1089.

\bibitem{9} A. Chodos, A.I. Hauser, and V. A. Kostelecky, Phys.
Lett. B {\bf 150} (1985) 295.

\bibitem{10} T. Chang, "Does a free tachyon exist ?", 
Proceedings of the Sir A. Eddington Centenary Symposium, Vol. 3, 
Gravitational Radiation and Relativity", p.431 (1986). 

\bibitem{11} A. Chodos, V.A. Kostelecky, R. Potting, 
and E. Gates, Mod. Phys. Lett. A {\bf 7} (1992) 467.

\bibitem{12} A. Chodos, and V.A. Kostelecky, Phys.Lett. B 
{\bf 336} (1994) 295.

\bibitem{13} E.C.G. Sudarshan: in Proceedings of the VIII 
Nobel Symposium, ed. by  N. Swartholm ( J. Wiley, New York, 1970), 
P.335;  

\bibitem{14} J. Bandukwala and D. Shay,  Phys. Rev. D {\bf 9} (1974) 
889;  D. Shay, Lett. Nuovo Cim. {\bf 19} (1977) 333. 

\bibitem{15}  ''Review of Particle Physics'', Euro. Phys. 
Journ. C {\bf 15} (2000) 350.

\bibitem{16} P.A.M. Dirac, Proc. R. Soc. Ser, A {\bf 117}; 610, 
{\bf 118} (1928) 351.

\bibitem{17} P.R. Holland, The Quantum Theory of Motion, 
Cambridge, UK, Chap. 12 (1993).

\bibitem{18} T.D. Lee and C.N. Yang, Phys. Rev. {\bf 104} 
(1956) 254; Phys. Rev. {\bf 105} (1957) 1671.

\bibitem{19} C.S. Wu et al., Phys. Rev. {\bf 105} (1957) 1413.

\bibitem{20} G-j. Ni et al, Chin. Phys. Lett., {\bf 17} (2000) 393.

\bibitem{21} T. Chang and G-j. Ni, "An Explanation on Negative
Mass-Square of Neutrinos", LANL preprint hep-ph/0009291 (2000);
 G-j. Ni and T. Chang, "Is Neutrino a Superluminal Particle ?", 
hep-ph/0103051 (2001). 

\bibitem{22} A. Einstein, H.A. Lorentz, H. Minkowski, 
and H. Weyl, The Principle of Relativity (collected papers), 
Dover, New York (1952).

\bibitem{23} R. Tangherlini, Nuov. Cim Suppl., {\bf 20} (1961) 1.

\bibitem{24} T. Chang, J. Phys. A {\bf 12} (1979) L203. 

\bibitem{25} J. Rembielinski, Phys. Lett., A {\bf 78} (1980) 33;
Int. J. Mod. Phys.,A {\bf 12} (1997) 1677.

\bibitem{26} T. Chang and D.G. Torr, Found. Phys. Lett., 
{\bf 1} (1988) 343.

\bibitem{27} J. Ciborowski and J. Rembielinski, Europ. Phys. J., C 
 {\bf 8} (1999) 157.

\end{thebibliography}
\end{document}